\documentclass[11pt,a4paper]{article}

\usepackage{amsmath,amsfonts,graphicx,natbib,psfrag,color}
\usepackage{algorithm}
\usepackage[textsize=small]{todonotes}
\usepackage{booktabs}
\graphicspath{{graphics/}}

\title{Correlated pseudo-marginal schemes for time-discretised stochastic kinetic models}
\author{Andrew Golightly$^{a,}$\thanks{email:
    \texttt{andrew.golightly@newcastle.ac.uk}} \and\ Emma Bradley$^b$ \and\ Tom Lowe$^a$
  \and\ Colin S. Gillespie$^a$}
\date{$^a$ School of Mathematics, Statistics and Physics, Newcastle University,
  Newcastle upon Tyne, UK\\
$^b$ Hiscox Ltd., London, UK}

\begin{document}
\maketitle
\begin{abstract}
The challenging problem of conducting fully Bayesian inference for the reaction rate constants governing 
stochastic kinetic models (SKMs) is considered. Given the challenges underlying this problem, 
the Markov jump process representation is routinely replaced by an approximation based on a suitable time-discretisation 
of the system of interest. Improving the accuracy of these schemes amounts to using an ever 
finer discretisation level, which in the context of the inference problem, requires integrating over the uncertainty in the process 
at a predetermined number of intermediate times between observations. Pseudo-marginal Metropolis--Hastings schemes are increasingly used, 
since for a given discretisation level, the observed data likelihood can be unbiasedly estimated using a particle filter. When 
observations are particularly informative, an auxiliary particle filter can be implemented, by propagating particles conditional 
on the next observation. Recent work in state-space 
settings has shown how the pseudo-marginal approach can be made much more efficient by correlating the underlying 
pseudo-random numbers used to form the estimate of likelihood at the current and proposed values of the unknown parameters. 
This approach is extended to the time discretised SKM framework by correlating the innovations that drive the auxiliary 
particle filter. The resulting approach is found to offer substantial gains in efficiency over a standard 
implementation.    

\end{abstract}


\section{Introduction}
\label{sec:intro}

A stochastic kinetic model (SKM) typically refers to a reaction network, an associated rate law and a probabilistic 
description of the reaction dynamics. Reactions occur continuously in time with a reaction occurrence resulting in 
a discrete change to the system state. A Markov jump process (MJP) therefore provides a natural description of the time-course 
behaviour of the species involved in the reaction network. The resulting modelling framework is fairly flexible 
and consequently, has been used ubiquitously in areas such as epidemiology \citep{lin2013b,mckinley2014,oneill1999}, 
population ecology \citep{ferm2008,BWK08,Gillespie2010} and systems biology \citep{Wilkinson09,GoliWilk15,koblents2015,Hey2015}. 
A concise introduction to SKMs can be found in \cite{Wilkinson06}.

Whilst exact simulation of the MJP is straightforward (using for example the direct method of \cite{Gillespie1976}), 
performing exact fully Bayesian inference is made problematic by the intractability of the observed data likelihood. 
Consequently, several approaches have been developed that make use of computationally intensive methods. 
These include the use of data augmentation \citep{boys2007,BWK08} together with Markov chain Monte Carlo (MCMC), 
reversible jump MCMC \citep{BWK08,Wang2010}, population Monte Carlo 
\citep{koblents2015} and particle MCMC \citep{andrieu10,GoliWilk11,Owen:2015}. Such methods typically require many 
simulations of the jump process, prohibiting their use for SKMs with many reactions and species. Consequently, 
there has been much interest in the development of exact (simulation-based) inference schemes for cheap approximations 
of the MJP. In particular, approximations based on time-discretisation do not require simulation of every reaction event, 
but rather update the state of the system in one go, after a particular time-step (typically chosen by the practitioner). 
In this paper, we focus on two such approximations, the Poisson leap \citep{Gillespie01,Anderson2008} and chemical Langevin equation, also known as the 
diffusion approximation \citep{Gillespie92b,Gillespie00,GoliWilk05}. Our umbrella aim is the development of fully Bayesian inference schemes that are 
both computationally and statistically efficient. 

When working with the time-discretised process, inference is still far from straightforward. Ensuring a desired level 
of accuracy requires the introduction of a pre-specified number of intermediate time-points between observations. Since 
the latent process at these time-points cannot be integrated out analytically, the observed data likelihood under the 
approximate model remains intractable. We therefore develop a particle MCMC scheme for performing fully Bayesian 
inference for either the Poisson leap or CLE and improve computational efficiency over a vanilla implementation in two ways. 
First, an auxiliary particle filter \citep{pitt1999} is used to (unbiasedly) estimate the observed data likelihood. As shown by 
\cite{GoliWilk15}, this is crucial in avoiding highly variable likelihood estimates in scenarios where intrinsic stochasticity 
outweighs the error in the observation process. Essentially, particles are propagated conditional on future observations 
by using a suitable bridge construct. When using the Poisson leap, we propose to use the conditioned reaction hazard 
of \cite{GoliWilk15}. For the CLE, we use the modified diffusion bridge (MDB) of \cite{DurhGall02} (or the appropriate extension to 
incomplete observation described in \cite{whitaker2017}). Finally, we make use of the recently proposed correlated 
pseudo-marginal algorithm \citep{deligiannidis2016,dahlin2015}, which introduces positive correlation between successive likelihood 
estimates in order to reduce the variance of the acceptance ratio. 

Our approach is to introduce correlation between the 
bridges generated by the auxiliary particle filter at iteration $i$ and those generated at iteration $i+1$. When using the CLE, 
this can be achieved by correlating the Gaussian innovations that drive the MDB. When using the Poisson leap, the numbers of 
reaction events conditional on the next observation can be used. A similar approach is described in \cite{tran2016} for a univariate 
diffusion process. \cite{choppala2016} consider a Lotka--Volterra reaction network and calculate the observed data likelihood 
by averaging $G$ `blocks' of unbiased estimates (which can be computed in parallel). Correlation is introduced by only updating 
the likelihood in a randomly chosen block. This is the so-called blockwise pseudo-marginal method. Our contribution is a 
unified framework for applying a correlated pseudo-marginal algorithm to a general class of time-discretised stochastic kinetic 
models, that additionally allows a flexible observation regime. In particular, we consider incomplete observation of the model 
components as well as Gaussian measurement error. We apply the resulting methodology to four examples arising in systems biology 
and epidemiology, using both real and synthetic data. 

The remainder of this paper is organised as follows. Section~\ref{sec:stochkin} gives a brief introduction to SKMs with particular 
attention to the derivation of the Poisson leap and CLE approximations. The inference algorithm is described in detail in 
Section~\ref{sec:bayes} and applied to several examples in Section~\ref{app}. Conclusions are drawn in Section~\ref{conc}.

\section{Stochastic kinetic models}
\label{sec:stochkin}

Consider a reaction network involving $s$ species $\mathcal{X}_1,\ldots,\mathcal{X}_s$ and $v$ reactions $\mathcal{R}_1,\ldots,\mathcal{R}_v$ 
such that
\[
 \sum_{j=1}^{s}p_{ij}\mathcal{X}_j \longrightarrow
\sum_{j=1}^{s}q_{ij}\mathcal{X}_j, \quad i=1,\ldots,v 
\]
where $p_{ij}$ and $q_{ij}$ are non-negative integers known as stoichiometric coefficients. 
Let $X_{j,t}$ denote the (discrete) number of species $\mathcal{X}_j$ at time
$t$, and let $X_t$ be the $s$-vector $X_t = (X_{1,t},\linebreak[1] \ldots,\linebreak[0] X_{s,t})^T$. 
The time evolution of $X_t$ is most naturally described by a Markov jump process (MJP), 
so that for an infinitesimal time increment $dt$ and an instantaneous hazard $h_i(X_t,c_i)$, 
the probability of a type $i$ reaction occurring in the time interval $(t,t+dt]$ is $h_i(X_t,c_i)dt$. 
Under the standard assumption of mass action kinetics, $h_i$ is proportional to a product of 
binomial coefficients. Specifically
\[
h_i(X_t,c_i) = c_i\prod_{j=1}^s \binom{X_{j,t}}{p_{ij}}.
\] 
Values for $c=(c_1,\ldots,c_v)^T$, the initial system state $X_0=x_0$ and 
the $s\times v$ stoichiometry matrix $S$ whose $(i,j)$th element is given by $q_{ji}-p_{ji}$, 
complete specification of the Markov process. Despite the intractability of the probability 
mass function governing the state of the system at any time $t$, generating exact realisations 
of the MJP is straightforward via a technique known in this context as 
\emph{Gillespie's direct method} \citep{Gillespie77}. In brief, if the current time and state of the system are $t$ and
$X_t$ respectively, then the time to the next event will be
exponentially distributed with rate parameter
\[
h_0(X_t,c)=\sum_{i=1}^v h_i(X_t,c_i),
\]
and the event will be a reaction of type $\mathcal{R}_i$ with probability
$h_i(X_t,c_i)/h_0(X_t,c)$ independently of the inter-event time.

\subsection{Time-discretisation}
\label{sec:disc}

Whilst generating simulations of the MJP description of the SKM is straightforward, 
capturing every occurrence of a reaction time and type can be computationally expensive, 
and this may preclude use of the MJP as an inferential model. We therefore consider 
two approximations to the MJP, the Poisson leap method and the chemical Langevin equation, 
and give a brief, informal derivation of both approaches.

Consider an infinitesimal time interval, $(t,t+dt]$, over which the
reaction hazards will remain constant almost surely. The occurrence of reaction events can 
therefore be regarded as the occurrence of events of a Poisson process with independent realisations
for each reaction type. Hence, for an interval $(t,t+\Delta t]$ of finite length, $\Delta t$, and the 
current system state $X_t$, the number of reaction events of type $i$, $r_i$, is Poisson distributed 
with rate $h_i(X_t,c)\Delta t$. Let $r=(r_1,\ldots,r_v)^T$. It should then be clear that the 
system state can be updated approximately, according to
\begin{equation}\label{pl}
X_{t+\Delta t}=X_t+Sr\,.
\end{equation}  
Further extensions to this approach (although not pursued here) involve stepping 
ahead a variable amount of time $\tau$, based on the rate constants 
and the current state of the system. This gives the so called 
$\tau$-leap algorithm \citep{Gillespie01}.

It should now be clear from (\ref{pl}) that the expectation and variance of 
the infinitesimal $d X_t$ are
\[
\operatorname{E}(dX_t)=S\,h(X_t,c)dt,\qquad \operatorname{Var}(dX_t)= S\operatorname{diag}\{h(X_t,c)\}S^Tdt,
\]
where $h(X_t,c)=(h_1(X_t,c_1),\ldots,h_v(X_t,c_v))^T$. Hence, a further approximation can be obtained by 
constructing the It\^o stochastic differential equation (SDE) that has the same infinitesimal mean 
and variance as the true MJP. That is
\begin{equation}
dX_t = S\,h(X_t,c)dt + \sqrt{S\operatorname{diag}\{h(X_t,c)\}S^T}\,dW_t,
\label{cle}
\end{equation}
where $W_t$ is an $s$-vector of standard Brownian 
motion and $\sqrt{S\operatorname{diag}\{h(X_t,c)\}S^T}$ 
is an $s\times s$ matrix $B$ such that $BB^T=S\operatorname{diag}\{h(X_t,c)\}S^T$. 
Equation \eqref{cle}
is the SDE most commonly referred to as the chemical Langevin
equation (CLE), and can be shown to approximate the SKM increasingly well in high
concentration scenarios \citep{Gillespie00}. The CLE can rarely be solved analytically, 
and it is common to work with a discretisation such as the Euler--Maruyama discretisation which gives
\[
X_{t+\Delta t} = X_t+ S\,h(X_t,c)\Delta t + \sqrt{S\operatorname{diag}\{h(X_t,c)\}S^T\Delta t}\,Z
\]
where $Z$ is a standard multivariate Gaussian random variable.

\section{Bayesian inference}
\label{sec:bayes}

\subsection{Setup}
\label{sec:setup}

Suppose that the Markov jump process is not observed directly, 
but observations $y_{t}$ (on a regular grid) are available 
and assumed conditionally independent (given the latent jump process) with conditional probability 
distribution obtained via the observation equation,
\begin{equation}\label{obs_eq}
Y_{t}=P^TX_{t}+\varepsilon_{t},\qquad \varepsilon_{t}\sim \textrm{N}\left(0,\Sigma\right),\qquad t=1,\ldots, n.
\end{equation} 
Here, $Y_{t}$ is a length-$p$ vector, $P$ is a constant matrix of dimension 
$s\times p$ and $\varepsilon_{t}$ is a length-$p$ 
Gaussian random vector. The density linking the observed and latent process is denoted by 
$p(y_{t}|x_{t})$. For simplicity we assume that $\Sigma$ is known.

In what follows, we replace the expensive MJP with either the Poisson leap 
approximation or chemical Langevin equation, and perform exact (simulation-based) Bayesian 
inference using the approximate model. We anticipate that the time between observations 
is too large for these approximations to be directly applied and therefore introduce 
intermediate times between observations. Hence, without loss of generality, 
consider an equally spaced partition of the time interval $[t-1,t]$ 
as
\begin{equation}\label{disc}
t-1=\tau_{t-1,0}<\tau_{t-1,1}<\ldots <\tau_{t-1,m-1}<\tau_{t-1,m}=t
\end{equation} 
where $\tau_{t-1,i+1}-\tau_{t-1,i}=\Delta\tau=1/m$. Hence, the approximation is applied recursively 
over each sub-interval $[\tau_{t-1,i},\tau_{t-1,i+1}]$ rather than in a single instance over 
$[t-1,t]$. Note that the value $m$ is pre-specified by the 
practitioner and controls both the accuracy and computational cost of the approximation. 

Suppose now that the main objective is inference for the rate constants $c$ given data 
$y=(y_1,\ldots,y_n)^T$. To this end, consider the marginal posterior 
\begin{equation}\label{post}
\pi(c)\propto \pi_0(c)p(y|c)
\end{equation}
where $\pi_0(c)$ is the prior density ascribed to $c$. Unfortunately, (\ref{post}) is 
complicated by the observed data likelihood $p(y|c)$. For the CLE, this satisfies
\[
p(y|c)=\int p(x|c)p(y|x)dx
\]
where $x=(x_{\tau_{1,0}},\ldots,x_{\tau_{1,m}},x_{\tau_{2,0}},x_{\tau_{2,1}}\ldots\ldots,x_{\tau_{n-1,m}})$. 
Additionally,
\begin{equation}\label{xdens}
p(x|c)=p(x_1)\prod_{t=1}^{n-1}\prod_{i=0}^{m-1}\textrm{N}\left(x_{\tau_{t,i+1}};x_{\tau_{t,i}}+ S\,h(x_{\tau_{t,i}},c)\Delta \tau\,,\, S\operatorname{diag}\{h(x_{\tau_{t,i}},c)\}S^T\Delta \tau\right)
\end{equation}
and
\begin{equation}\label{obsdens}
p(y|x)=\prod_{t=1}^{n}\textrm{N}\left(y_t;P^T x_t\,,\,\Sigma\right)
\end{equation}  
where $\textrm{N}(\cdot;a,B)$ denotes the pdf of a Gaussian random variable with mean $a$ and variance $B$. For the Poisson leap 
approximation we have that
\[
p(y|c)=\sum_{x_{1},r} p(x_1)p(r|x_{1},c)p(y|r,x_1)
\] 
where $r=(r_{\tau_{1,1}},\ldots,r_{\tau_{1,m}},r_{\tau_{2,1}},r_{\tau_{2,2}}\ldots\ldots,r_{\tau_{n-1,m}})$ and for example, $r_{\tau_{t,i}}=(r_{\tau_{t,i,1}},\ldots,r_{\tau_{t,i,v}})^T$ is the 
length-$v$ vector containing the number of reactions of each type in the interval $[\tau_{t,i-1},\tau_{t,i}]$. It should be clear that given $x_1$ and $r$, 
$x$ can be obtained deterministically through recursive application of (\ref{pl}). Hence $p(y|r,x_1)$ coincides with $p(y|x)$ above and
\[
p(r|x_{1},c)=\prod_{t=1}^{n-1}\prod_{i=0}^{m-1}\prod_{j=1}^{v}\textrm{Po}\left(r_{\tau_{t,i+1,j}}\,;\,h_{j}(x_{\tau_{t,i}},c_j)\Delta\tau\right)
\]
where $\textrm{Po}(\cdot;h)$ denotes the mass function of a Poisson random variable with mean $h$.

Owing to the intractability of $p(y|c)$, the posterior in (\ref{post}) 
is sampled via Markov chain Monte Carlo (MCMC). In particular, a suitably 
constructed pseudo-marginal Metropolis--Hastings scheme (PMMH) \citep{beaumont03,andrieu09b,andrieu10} 
provides an effective way of performing this task. We briefly describe this approach in the 
next section alongside an adaptation of a recently proposed modification (the so-called 
correlated PMMH method) that gives a significant improvement in efficiency over the basic scheme.


\subsection{Correlated pseudo-marginal Metropolis--Hastings}
\label{sec:corpmmh}

Suppose that auxiliary variables $U\sim g(u)$ can be used to generate 
a non-negative unbiased estimator $\hat{p}_{U}({y}|c)$ of $p({y}|c)$. 
Therefore, an unbiased (up to a constant) estimator of the posterior is
\[
\hat{\pi}_{U}(c) = \pi_{0}(c) \hat{p}_{U}({y}|c).
\]
The PMMH scheme is an MH scheme targeting
\[
\tilde{\pi}(c,u)=\pi_{0}(c)g(u)\hat{p}_{u}(y|c)
\]
which has marginal distribution
\[
\int \pi_{0}(c)g(u)\hat{p}_{u}(y|c)\, du \propto \pi(c).
\]
For a proposal kernel of the form $q(c'|c)g(u')$, the MH acceptance probability 
is
\begin{align}
\alpha\left\{(c',u')|(c,u) \right\}&=\textrm{min}\left\{1\,,\,\frac{\tilde{\pi}(c',u')}{\tilde{\pi}(c,u)}\times\frac{q(c|c')g(u)}{q(c'|c)g(u')}  \right\}\nonumber \\
&=\textrm{min}\left\{1\,,\, \frac{\pi_{0}(c')\hat{p}_{u'}(y|c')}{\pi_{0}(c)\hat{p}_{u}(y|c)}\times \frac{q(c|c')}{q(c'|c)}\right\}\label{alpha}
\end{align}
and therefore the density associated with the auxiliary variables need not be evaluated. 

Note that the proposal kernel need not be restricted to the use of $g(u')$. The correlated PMMH scheme \citep{deligiannidis2016,dahlin2015} generalises the PMMH scheme by using a proposal kernel of the form $q(c'|c)K(u'|u)$ where $K(\cdot|\cdot)$ satisfies the detailed balance equation
\begin{equation}\label{Kdb}
g(u)K(u'|u)=g(u')K(u|u').
\end{equation}
It is straightforward to show that a MH scheme with proposal kernel $q(c'|c)K(u'|u)$ and acceptance probability (\ref{alpha}) 
satisfies detailed balance with respect to the target $\tilde{\pi}(c,u)$. Upon negating the trivial scenario that the chain does not move, 
we have that
\begin{align*} 
&\tilde{\pi}(c,u)q(c'|c)K(u'|u)\alpha\left\{(c',u')|(c,u) \right\}\\
&\qquad =\textrm{min}\left\{\pi_{0}(c)g(u)\hat{p}_{u}(y|c)q(c'|c)K(u'|u)\,,\,\pi_{0}(c')g(u)\hat{p}_{u'}(y|c')q(c|c')K(u'|u)\right\}\\
&\qquad =\textrm{min}\left\{\pi_{0}(c)g(u)\hat{p}_{u}(y|c)q(c'|c)K(u'|u)\,,\,\pi_{0}(c')g(u')\hat{p}_{u'}(y|c')q(c|c')K(u|u')\right\}\\
&\qquad =\tilde{\pi}(c',u')q(c|c')K(u|u')\alpha\left\{(c,u)|(c',u') \right\}
\end{align*}
where (\ref{Kdb}) is used to deduce the third line. 

In practice, $g(u)$ is a standard Gaussian density and $K(u'|u)$ is taken to be the kernel associated with a Crank--Nicolson proposal. That is
\[
g(u)=\textrm{N}\left(u;\,0\,,\,I_d\right)\qquad \textrm{and} \qquad K(u'|u)=\textrm{N}\left(u';\,\rho u\,,\,\left(1-\rho^2\right)I_d\right)
\]
where $I_d$ is the identity matrix whose dimension $d$ is determined by the number of elements in $u$ and $\rho$ is chosen 
to be close to 1, to induce positive correlation between $\hat{p}_{U}({y}|c)$ and $\hat{p}_{U'}({y}|c')$. Taking $\rho=0$ 
gives the special case that $K(u'|u)=g(u')$, which corresponds to the PMMH scheme. The motivation for taking 
$\rho\approx 1$ is to reduce the variance of the acceptance probability in (\ref{alpha}). Consequently, significant gains in statistical 
efficiency (relative to the standard PMMH scheme) may be expected. In scenarios where $U$ is not normally distributed (as is the case for the Poisson leap approximation) it is 
straightforward to generate uniform random variates via $\Phi(U)$ (where $\Phi(\cdot)$ is the cdf of a standard normal random 
variable). These uniform draws can then be transformed e.g. to give Poisson draws, via the inversion method. 

The correlated PMMH scheme is summarised in Algorithm~\ref{algcor}. After initialisation, each iteration requires computation of 
$\hat{p}_{u'}({y}|c')$. This is achieved by executing an auxiliary particle filter (for each proposed value $(c',u')$), 
which we describe in the next section. 
\begin{algorithm}[t]
\caption{Correlated PMMH scheme (CPMMH)}\label{algcor}
\textbf{Input:} correlation parameter $\rho$ and the number of CPMMH iterations $n_{\textrm{iters}}$.\\
\textbf{Output:} $c^{(1)},\ldots,c^{(n_{\textrm{iters}})}$.
\begin{enumerate}
\item For iteration $i=0$:
\begin{itemize}
\item[(a)] Set $c^{(0)}$ in the support of $\pi(c)$ and draw $u^{(0)}\sim \textrm{N}(0,I_d)$.
\item[(b)] Compute $\hat{p}_{u^{(0)}}({y}|c^{(0)})$ by running Algorithm~\ref{auxPF} with $(c,u)=(c^{(0)},u^{(0)})$.
\end{itemize}
\item For iteration $i=1,\ldots, n_{\textrm{iters}}$:
\begin{itemize}
\item[(a)] Draw $c'\sim q(\cdot | c^{(i-1)})$ and $\omega\sim \textrm{N}(0,I_d)$. Put $u'=\rho u^{(i-1)}+\sqrt{1-\rho^2}\omega$.
\item[(b)] Compute $\hat{p}_{u'}({y}|c')$ by running Algorithm~\ref{auxPF} with $(c,u)=(c',u')$.
\item[(c)] With probability $\alpha\left\{(c',u')|(c^{(i-1)},u^{(i-1)}) \right\}$ given by (\ref{alpha}), 
put $(c^{(i)},u^{(i)})=(c',u')$ otherwise store the current values $(c^{(i)},u^{(i)})=(c^{(i-1)},u^{(i-1)})$.
\end{itemize}
\end{enumerate}
\end{algorithm}

\subsection{Auxiliary particle filter}
\label{sec:aux}
   
The observed data likelihood $p(y|c)$ can be factorised as 
\begin{equation}\label{margllfact}
p(y|c)=p(y_{1}|c)\prod_{t=2}^{n}p(y_{t}|y_{1:t-1},c)
\end{equation}
where $y_{1:t-1}=(y_1,\ldots,y_{t-1})$. Although the constituent terms in (\ref{margllfact}) 
will typically be intractable, a particle filter provides an efficient mechanism 
for their estimation. Moreover, the particle filter that we consider here gives 
an unbiased estimator of $p(y|c)$ \citep{delmoral04,pitt12} and hence can be used in steps 1(b) and 2(b) 
of the CPMMH scheme; see Algorithm~\ref{algcor}. For a concise introduction to particle filters, 
we refer the reader to \cite{Kunsch2013} and the references therein. 

The basic idea behind the particle filter is to recursively approximate the sequence of filtering 
densities $p(x_{t}|y_{1:t},c)$ using a sequence of importance sampling and resampling steps. Let $u=(u_1,\ldots,u_n)$ 
denote a realisation of the random variables required by the particle filter. We further adopt the partition 
$u_t=(\tilde{u}_t,\bar{u}_t)^T$ to distinguish between the variables used to propagate state particles and those 
used in the resampling step, respectively. Note that $\tilde{u}_t=(\tilde{u}_t^1,\ldots,\tilde{u}_t^N)$ corresponding 
to a filter with $N$ particles and $\tilde{u}_t^i=(\tilde{u}_{t,1}^i,\ldots,\tilde{u}_{t,m}^i)$ for $t>1$, corresponding to 
the discretisation introduced in Section~\ref{sec:setup}. 

Given a weighted sample of `particles' $\{x_{t-1}^i,w(u_{t-1}^i)\}_{i=1}^{N}$ approximately 
distributed according to $p(x_{t-1}|y_{1:t-1},c)$, the particle filter uses the approximation
\begin{equation}\label{pftarget}
\hat{p}(x_{(t-1,t]}|y_{1:t},c)\propto p(y_{t}|x_{t},c)\sum_{i=1}^{N}p(x_{(t-1,t]}|x_{t-1}^{i},c)w(u_{t-1}^i)
\end{equation}
where, in the case of the CLE, $x_{(t-1,t]}=(x_{\tau_{t-1,1}},\ldots,x_{\tau_{t-1,m}})$. We focus here on the CLE 
for reasons of brevity but note that in the case of the Poisson leap approximation, $x_{(t-1,t]}$ is replaced 
by $r_{(t-1,t]}=(r_{\tau_{t-1,1}},\ldots,r_{\tau_{t-1,m}})$ since $x_t$ can be obtained deterministically using 
the state $x_{t-1}$ and the number of reactions of each type in the interval $(t-1,t]$. 

The form of (\ref{pftarget}) 
suggests a simple importance sampling/resampling strategy where particles are resampled (with replacement) in proportion to their weights, 
propagated via $x_{(t-1,t]}^i=f_t(\tilde{u}_t^i)\sim p(\cdot|x_{t-1}^{i},c)$ and reweighted by $p(y_{t}|x_{t}^i,c)$. Here, 
$f_t(\cdot)$ is a deterministic function of $\tilde{u}_t^i$ (and the parameters $c$) that gives an explicit connection between 
the particles and auxiliary variables. Repeating this procedure at each time point gives the bootstrap particle filter (BPF) of \cite{gordon93}. 
As discussed in \cite{delmoral14} and \cite{GoliWilk15} however, this scheme is likely to perform poorly when observations 
are informative. In this case very few particles will have reasonable weight, leading to an estimator of observed data likelihood 
with high variance. This problem can be alleviated through the use of 
an auxiliary particle filter (APF) \citep{pitt1999,pitt12} 
which propagates particles via $x_{(t-1,t]}^i=f_t(\tilde{u}_t^i)\sim g(\cdot|x_{t-1}^i,y_{t},c)$, with the special 
case of $g(\cdot|x_{t-1}^i,y_{t},c)=p(\cdot|x_{t-1}^{i},c)$ giving the BPF. 

The APF is described generically in Algorithm~\ref{auxPF}. The output of the APF can be used to estimate 
the constituent terms in (\ref{margllfact}) by simply taking the average unnormalised weight; see steps 1(c) and 2(e). 
A discussion of the sorting and resampling steps 2(a) and 2(b) is provided in Section~\ref{sec:resamp}. Suitable propagation densities 
$g(x_{(t-1,t]}|x_{t-1},y_{t},c)$ and $g(r_{(t-1,t]}|x_{t-1},y_{t},c)$ (as necessary for the Poisson leap) are given in 
Section~\ref{sec:prop}.  

\begin{algorithm}[t]
\caption{Auxiliary particle filter}\label{auxPF}
\textbf{Input:} parameters $c$, auxiliary variables $u$ and the number of particles $N$.\\
\textbf{Output:} estimate $\hat{p}_{u}({y}|c)$ of the observed data likelihood.
\begin{enumerate}
\item Initialisation ($t=1$). 
\begin{itemize}
\item[(a)] \textbf{Sample} the prior. Draw $\tilde{u}_1^i\sim \textrm{N}(0,1)$ and put $x_{1}^{i}=f_1(\tilde{u}_1^i)\sim p(\cdot)$, $i=1,\ldots,N$.
\item[(b)] \textbf{Compute} the weights. For $i=1,\ldots,N$ set
\[
\tilde{w}(u_1^{i})=p(y_{1}|x_{1}^{i},c), \qquad w(u_1^{i})=\frac{\tilde{w}(u_1^i)}{\sum_{j=1}^{N}\tilde{w}(u_1^j)}.
\]
\item[(c)] \textbf{Update} observed data likelihood estimate. Compute $\hat{p}_{u_1}(y_{1}|c)=\sum_{i=1}^{N}\tilde{w}(u_1^i)/N$.
\end{itemize}
\item For times $t=2,3,\ldots ,n$:
\begin{itemize}
\item[(a)] \textbf{Sort.} Obtain the sorted index $s(i)$ and put
  $\left\{x_{t-1}^i,w(u_{t-1}^i)\right\}:=\left\{x_{t-1}^{s(i)},w(u_{t-1}^{s(i)})\right\}$,
  $i=1,\ldots,N$.
\item[(b)] \textbf{Resample.} Obtain ancestor indices $a_{t-1}^i$, $i=1,\ldots,N$ using systematic resampling on the collection of weights 
$\{w(u_{t-1}^1),\ldots,w(u_{t-1}^N)\}$.
\item[(c)] \textbf{Propagate.} Draw $\tilde{u}_t^i\sim \textrm{N}(0_{m},I_{m})$ and 
put $x_{(t-1,t]}^{i}=f_t(\tilde{u}_t^i)\sim g\big(\cdot|x_{t-1}^{a_{t-1}^{i}},y_{t},c\big)$, $i=1,\ldots,N$.
\item[(d)] \textbf{Compute} the weights. For $i=1,\ldots,N$ set
\[
\tilde{w}(u_t^i)=\frac{p(y_{t}|x_{t}^{i},c)p\big(x_{(t-1,t]}^{i}|x_{t-1}^{a_{t-1}^{i}},c\big)}
{g\big(x_{(t-1,t]}^{i}|x_{t-1}^{a_{t-1}^{i}},y_{t},c\big)}, \qquad w(u_t^{i})=\frac{\tilde{w}(u_t^i)}{\sum_{j=1}^{N}\tilde{w}(u_t^j)}.
\]
\item[(e)] \textbf{Update} observed data likelihood estimate. Compute
\[
\hat{p}_{u_{1:t}}(y_{1:t}|c)=\hat{p}_{u_{1:t-1}}(y_{1:t-1}|c)\hat{p}_{u_{t}}(y_{t}|y_{1:t-1},c)
\]
where $\hat{p}_{u_{t}}(y_{t}|y_{1:t-1},c)=\frac{1}{N}\sum_{i=1}^{N}\tilde{w}(u_t^i)$.

\end{itemize}
\end{enumerate}
\end{algorithm}

\subsubsection{Resampling}
\label{sec:resamp}

For the resampling step we follow \cite{deligiannidis2016} and use systematic resampling, which only requires 
simulating a single uniform random variable at each time point. These can be constructed from $\bar{u}_t\sim \textrm{N}(0,1)$ 
via $\Phi(\bar{u}_t)$. Sorted uniforms can then be found via $\bar{u}_{Rt}^i=(i-1)/N+\Phi(\bar{u}_t)/N, i=1,\ldots,N$ 
which are in turn used to choose indices $a_{t-1}^i$ that (marginally) satisfy $\textrm{Pr}(a_{t-1}^i=k)=w(u_{t-1}^k)$. 
Note that upon changing $c$ and $u$ the effect of the resampling step is likely to prune out different particles, thus 
breaking the correlation between successive estimates of observed data likelihood. To alleviate this problem, 
\cite{deligiannidis2016} sort the particles before resampling via the Hilbert sort procedure of \cite{gerber2015}. We follow 
\cite{choppala2016} by using a simple Euclidean sorting procedure. At observation time $t$ (immediately after propagation), 
we sort the particle trajectories $x_{(t-1,t]}^i$ as follows. The first sorted particle corresponds to that with the smallest 
value of the first component of the set $\{x_t^1,\ldots,x_t^N\}$. The remaining particles are chosen by minimising the Euclidean 
distance between the currently selected particle and the set of all other particles. 

\subsubsection{Propagation}
\label{sec:prop}
We require a suitable importance proposal 
$g(x_{(t-1,t]}|x_{t-1},y_{t},c)$ (or $g(r_{(t-1,t]}|x_{t-1},y_{t},c)$ in the case of the Poisson leap method) that takes into account 
the information in $y_t$. Consider a time interval $[t-1,t]$ and recall the partition in (\ref{disc}) which we will write as
\[
t-1=\tau_{0}<\tau_{1}<\ldots <\tau_{m-1}<\tau_{m}=t
\]
for notational simplicity. We adopt the following factorisations,
\[
g(x_{(t-1,t]}|x_{t-1},y_{t},c)=\prod_{k=0}^{m-1}g(x_{\tau_{k+1}}|x_{\tau_{k}},y_t,c),\qquad g(r_{(t-1,t]}|x_{t-1},y_{t},c)=\prod_{k=0}^{m-1}g(r_{\tau_{k+1}}|x_{\tau_{k}},y_t,c) 
\]
and seek suitable expressions for the constituent terms in each product. In the case of the CLE, we use the modified diffusion bridge construct of 
\cite{DurhGall02} (see also \cite{whitaker2017} for a recent discussion) which effectively uses a linear Gaussian approximation of $X_{\tau_{k+1}}|x_{\tau_{k}},y_t,c$. 
We obtain
\begin{equation}\label{imp1}
g(x_{\tau_{k+1}}|x_{\tau_{k}},y_t,c)=\textrm{N}\left(x_{\tau_{k+1}}\,;\, x_{\tau_k}+\mu(x_{\tau_k},c)\Delta\tau\,,\,\Psi(x_{\tau_k},c)\Delta\tau\right)
\end{equation}
where
\[
\mu(x_{\tau_k},c) =\alpha_{k}+\beta_{k} P\left(P^T\beta_{k}P\Delta_k + \Sigma\right)^{-1}
\left\{y_{t}-P^T(x_{\tau_{k}}+\alpha_{k}\Delta_k)\right\}   
\]
and
\[
\Psi(x_{\tau_k},c)=\beta_{k}-\beta_{k} P\left(P^T\beta_{k} P\Delta_k + \Sigma\right)^{-1}P^T\beta_{k}\Delta\tau. 
\]
Here $\alpha_{k}=S\,h(x_{\tau_k},c) $, $\beta_{k}=S\operatorname{diag}\{h(x_{\tau_k},c)\}S^T $ and $\Delta_{k}=t-\tau_{k}$. Given that the 
importance density in (\ref{imp1}) is Gaussian, it is straightforward to perform the propagation step in Algorithm~\ref{auxPF}. We draw 
$\tilde{u}_{t,k+1}^i\sim \textrm{N}(0,I_{s})$ and set
\[ 
x_{\tau_{k+1}}=x_{\tau_k}+\mu(x_{\tau_k},c)\Delta\tau+\sqrt{\Psi(x_{\tau_k},c)\Delta\tau}\,\tilde{u}_{t,k+1}^i,\qquad k=0,\ldots,m-1.
\]
For the Poisson leap approximation, we take $g(r_{\tau_{k+1}}|x_{\tau_{k}},y_t,c)$ to be a Poisson probability with rate given 
by an approximation to the expected number of reaction events in $[\tau_{k},\tau_{k+1}]$ given the current state of the system and, crucially, 
the observation $y_t$. The derivation of this approximate rate can be found in \cite{GoliWilk15} and is given by
\[
h^{*}(x_{\tau_k},c\,|y_{t})=h(x_{\tau_k},c)
+\operatorname{diag}\{h(x_{\tau_k},c)\} S^T P\left(P^T\beta_{k}P\Delta_k +\Sigma\right)^{-1}\left\{y_{t}-P^T(x_{\tau_{k}}+\alpha_{k}\Delta_k)\right\}. \label{haz}
\]
Hence, we obtain 
\begin{equation}\label{imp2}
g(r_{\tau_{k+1}}|x_{\tau_{k}},y_t,c)=\prod_{j=1}^{v}\textrm{Po}\left(r_{\tau_{k+1,j}}\,;\,h^{*}_j(x_{\tau_k},c\,|y_{t})\Delta\tau \right).
\end{equation}
The propagation step in Algorithm~\ref{auxPF} can be performed by first drawing $\tilde{u}_{t,k+1}^i\sim \textrm{N}(0,I_{v})$ and then 
applying the inverse Poisson CDF to each component of $\Phi(\tilde{u}_{t,k+1}^i)$ to give $r_{\tau_{k+1}}$, for $k=0,\ldots,m-1$. We then set
\[
x_{\tau_{k+1}}=x_{\tau_k}+S r_{\tau_{k+1}},\qquad k=0,\ldots,m-1.
\]

\subsection{Computational considerations}
\label{sec:comp}

A single iteration of the CPMMH scheme described in 
Algorithm~\ref{algcor} requires $n-1\times m\times N$ 
draws of the bridge construct with density (\ref{imp1}) 
when using the CLE, and mass function (\ref{imp2}) when 
using the Poisson leap. Recall that $n$ is the number 
of observations, $m$ is the number of latent process 
values per observation interval and $N$ is the number of 
particles in the auxiliary particle filter. The cost of drawing from 
(\ref{imp1}) and (\ref{imp2}) will be dictated by the number of 
observed components $p$, since the inversion of 
a $p\times p$ matrix is required. Nevertheless, for many systems of interest,
it is unlikely that all components will be observed \citep{GoliWilk15}, 
and we therefore anticipate that for systems with many species, $p << s$ where $s$ 
is the number of species. It remains for the practitioner 
to choose $m$ and $N$ to balance posterior accuracy and 
computational cost. 

We follow \cite{stramer11} and \cite{GoliWilk11} 
among others, by performing short pilot 
runs of the inference scheme (for a fixed, conservative value 
of $N$) with increasing values of $m$, until no 
discernible difference in the posterior output is detected (e.g. by visual 
inspection of kernel density estimates of the marginal parameter posteriors). For the 
examples in Section~\ref{app}, we find that $m \leq 10$ is 
sufficient.  

The number of particles $N$ controls the variance of the estimator of 
observed data likelihood $\hat{p}_{U}(y|c)$. As the variances increases, the acceptance 
probability of the pseudo-marginal MH scheme rapidly decreases to 0 \citep{pitt12}, 
resulting in `sticky' behaviour of the parameter chains. Practical advice for 
choosing $N$ to balance mixing performance and computational cost can found in 
\cite{doucet15} and \cite{sherlock2015}. The variance of the log-posterior 
(denoted $\sigma^{2}_{N}$, computed with $N$ particles) at a central value of $c$ 
(e.g. the estimated posterior median) should be around 2. For the CPMMH scheme, 
\cite{tran2016} suggests choosing $N$ so that $\sigma^{2}_{N}=2.16/1-\rho_{l}^2$ where 
$\rho_l$ is the estimated correlation between $\hat{p}_{u}(y|c)$ and $\hat{p}_{u'}(y|c')$. 
Note that for $\rho_l=0$ corresponding to the vanilla PMMH case, the aforementioned 
tuning advice is broadly consistent with \cite{doucet15} and \cite{sherlock2015}.

\section{Applications}\label{app}

To illustrate the proposed approach we consider four applications of increasing complexity. A simple immigration--death model 
is considered in Section~\ref{bd}. We fit the CLE to synthetic data and compare CPMMH with PMMH and additionally, the state-of-the-art 
MCMC scheme, that is, the modified innovation scheme (MIS) of \cite{GoliWilk08}, described briefly in the appendix. In Section~\ref{lv}, we fit the CLE associated 
with a Lotka--Volterra model to synthetic data. We also investigate the effect of increasing observation noise on the performance 
of the CPMMH scheme. The autoregulatory network of \cite{sherlock2014} is considered in Section~\ref{ar}. We generate synthetic data 
that is inherently discrete, precluding the use of the CLE as an inferential model. We therefore perform inference using the Poisson leap, 
and additionally explore the effect of using a bootstrap particle filter on the performance of the CPMMH scheme. Finally, the CLE approximation 
of a Susceptible--Infected--Removed (SIR) epidemic model is fitted using data on an influenza outbreak in a boys' boarding school in Great Britain 
\citep{bmj1978}. It is assumed that the infection rate is a mean reverting diffusion process giving a model with two unobserved components. 

Since the rate constants must be strictly positive we update 
$\log c$ using a random walk proposal with Gaussian innovations. We took the innovation variance to be the posterior 
variance of $\log c$ (estimated from a pilot run) scaled by a factor of $2.56^2/v$ for CPMMH and $2.38^2/v$ for MIS, 
where $v$ is the number of rate constants. We chose the number of particles $N$ by following the practical advice described 
in Section~\ref{sec:comp}. To ensure reasonable mixing of the auxiliary variables $U$, we adopted the conservative choice 
of $\rho=0.99$. In each example we use effective sample size (ESS) as a comparator. That is
\[
ESS=\frac{n_{\textrm{iters}}}{1+2\sum_{k=1}^{\infty}\alpha_k}
\]
where $\alpha_k$ is the autocorrelation function for the series at lag $k$ and $n_{\textrm{iters}}$ is the number 
of iterations in the main monitoring run. The ESS can be computed using the 
\verb+R+ package CODA \citep{Plummer06}. We report the minimum effective sample size over all components, denoted by ESS$_{\textrm{min}}$. 
All algorithms are coded in R and were run on a desktop computer with an Intel Core i7-4770 processor at 3.40GHz. An R implementation 
of PMMH, CPMMH and MIS for generic (univariate) diffusion processes is available at 
\emph{https://github.com/csgillespie/cor-pseudo-marginal}

\subsection{Immigration--death model}\label{bd}
The immigration--death reaction network takes the form
\[
\mathcal{R}_{1}:\, \emptyset \xrightarrow{\phantom{a}c_{1}\phantom{a}} \mathcal{X}_{1}\qquad
\mathcal{R}_{2}:\,\mathcal{X}_{1} \xrightarrow{\phantom{a}c_{2}\phantom{a}} \emptyset
\]
with immigration and death reactions shown respectively. The stoichiometry matrix is
\[
S = \begin{pmatrix} 
1 & -1 
\end{pmatrix}
\]
and the associated hazard function is 
\[
h(X_t,c) = (c_{1}, c_{2}X_{t})^T
\]
where $X_t$ denotes the state of the system at time $t$. Applying (\ref{cle}) directly gives the CLE as
\[
dX_{t}=\left(c_{1}-c_{2}X_{t}\right)\,dt + \sqrt{\left(c_{1}+c_{2}X_{t}\right)}\,dW_{t}.
\]
We generated a synthetic data set consisting of 101 observations by simulating from the 
Markov jump process via Gillespie's direct method and retaining the system state at 
integer times. To provide a challenging scenario for the CLE, we took $c_1=4$ and $c_2=0.8$ 
giving inherently discrete trajectories that `mean revert' around the value 5. Moreover, 
we took $X_0=500$ so that typical trajectories exhibit nonlinear dynamics over the time interval 
$[0,10]$. We assume error-free observation of $X_t$ so that the latent path between observation times, 
which is propagated according to equation (\ref{imp1}), becomes
\[
g(x_{\tau_{k+1}}|x_{\tau_{k}},x_t,c)=\textrm{N}\left(x_{\tau_{k+1}}\,;\, x_{\tau_k}+\frac{x_{t}-x_{\tau_{k}}}{t-\tau_{k}}\Delta\tau\,,\, 
\frac{t-\tau_{k+1}}{t-\tau_{k}}\beta(x_{\tau_{k}},c)\Delta\tau \right),
\] 
which can be sampled for $k=0,1,\ldots,m-2$. We also note in the case of error-free observation of all 
components of $X_t$ (as is considered in this application), the auxiliary particle filter can be seen 
as a simple importance sampler. Consequently, the sorting and resampling steps of Algorithm~\ref{auxPF} 
are not required here. 

\begin{figure}[t]
\centering
\psfragscanon
\psfrag{Density}[][][0.7][0]{Density}
\psfrag{log(c1)}[][][0.7][0]{$\log c_1$}
\psfrag{log(c2)}[][][0.7][0]{$\log c_2$}
\includegraphics[width=0.34\textwidth,angle=-90]{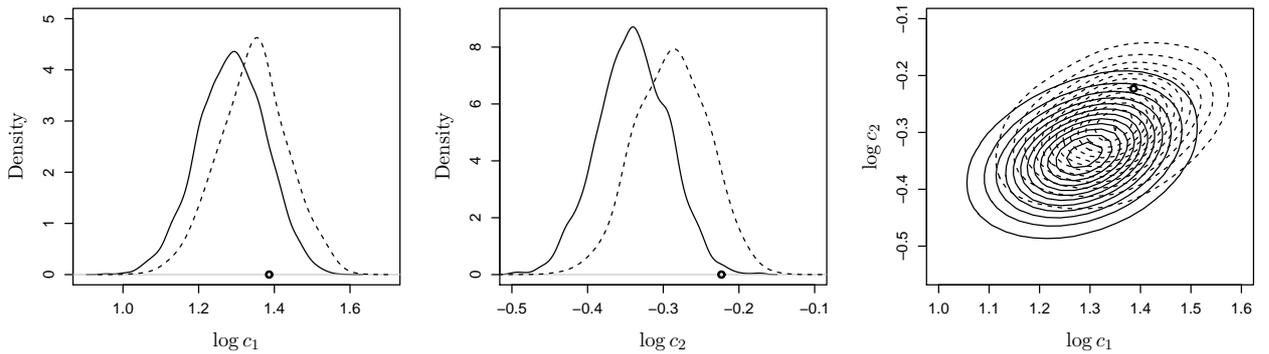}
\caption{Immigration--death model. Left and middle panels: marginal posterior distributions based on the CLE (solid lines) and MJP (dashed lines). Right panel: Contour plot of the joint posterior based on the CLE (solid line) and MJP (dashed line). The true values of $\log c_1$ and $\log c_2$ are indicated.}
\label{fig:figID}
\end{figure}

\begin{figure}[t]
\centering
\psfragscanon
\psfrag{Lag}[][][0.7][0]{Lag}
\includegraphics[width=0.34\textwidth,angle=-90]{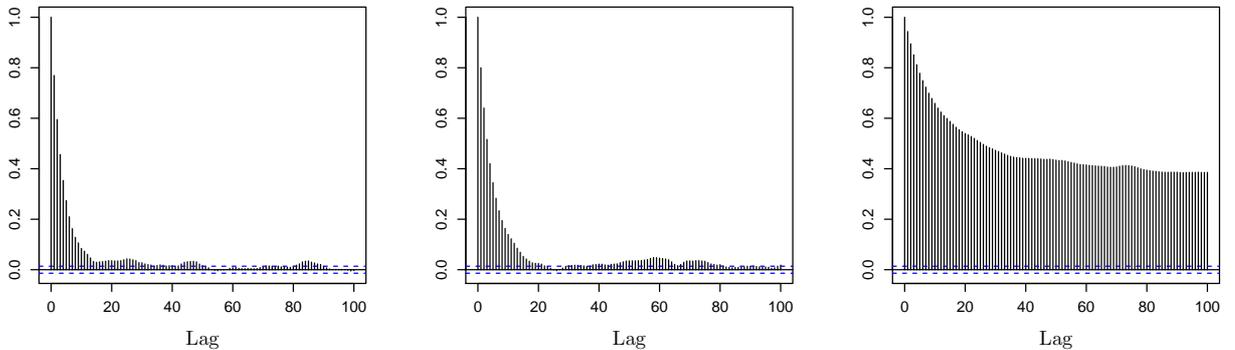}
\caption{Immigration--death model. Correlogram based on $\log c_2$ samples from the output of MIS (left panel), CPMMH with $\rho=0.99$ (middle panel) and 
PMMH (right panel).}
\label{fig:figID2}
\end{figure}

We took independent $N(0,10^2)$ priors for $\log c_1$ and $\log c_2$, and 
determined an appropriate discretisation level by performing short runs of MIS 
with $\Delta\tau\in\{0.05,0.1,0.2,0.5\}$. Since there was very little difference in posteriors 
beyond $\Delta\tau=0.2$, we used this value in the main monitoring runs which consisted of 
$2\times 10^4$ iterations of MIS, CPMMH and PMMH. The results are summarised by 
Figures~\ref{fig:figID}--\ref{fig:figID2} and Table~\ref{tab:tabID}.

Table~\ref{tab:tabID} shows a comparison of each 
competing inference scheme. Practical advice (as described above) suggests that CPMMH can tolerate much smaller values of $N$, 
with the scheme only requiring a value of $N$ around 2 (and we report results for $N=1,2$) when $\rho=0.99$ 
compared to $N=50$ for PMMH. Moreover, we found that the PMMH scheme often 
exhibited `sticky' behaviour, resulting in relatively low effective sample sizes. Consequently, in terms of 
minimum ESS per second, CPMMH (with $\rho=0.99$, $N=1$) outperforms PMMH by a factor of 210, reducing to 150 
when $N=2$.

As noted by \cite{deligiannidis2016}, 
values of $\rho$ close to 1 can result in slow mixing of the auxiliary variables $U$, in turn giving 
parameter correlograms that exhibit long range dependence. This does not appear to be the case for $\rho=0.99$ 
(see middle panel of Figure~\ref{fig:figID2}). Nevertheless, we note that reducing $\rho$ to 0.9 still gives 
an increase in overall efficiency of almost two orders of magnitude over PMMH. When comparing CPMMH to the modified 
innovation scheme we obtain similar ESS values. However, the relatively low 
computational cost of CPMMH (with $\rho=0.99$, $N=1$) results in an improvement in overall efficiency (with an mESS/s of 
42 vs 18). 
\begin{table}[t]
  \centering
  \small
  \begin{tabular}{@{}lllllll@{}}
    \toprule
    Algorithm  & $\rho$ &  $N$ & CPU (s) & mESS & mESS/s & Rel.  \\
    \midrule
    MIS        & \phantom{0.0}-- &\phantom{0}--& \phantom{0}121 & 2190 & 18     & \phantom{0}90\\
    CPMMH      & 0.99            &\phantom{0}1 & \phantom{00}45 & 1910 & 42     &   210 \\
               & 0.99           &\phantom{0}2 & \phantom{00}78  & 2370          & 30 & 150\\
               & 0.90           &\phantom{0}1 & \phantom{00}45  & \phantom{0}820& 18 & \phantom{0}90\\
    PMMH  & \phantom{.00}0       & 50          & 1740           &\phantom{0}380 & \phantom{0}0.2 &\phantom{00}1 \\
    \bottomrule
  \end{tabular}
  \caption{Immigration--death model. Correlation parameter $\rho$, number of particles $N$, 
    CPU time (in seconds $s$), minimum ESS, minimum ESS per second and relative (to PMMH) minimum ESS per second. All 
    results are based on $2\times 10^4$ iterations of each scheme.}\label{tab:tabID}	
\end{table}


The effect of the CLE as an inferential model can be seen in Figure~\ref{fig:figID}. Marginal posteriors based on the CLE 
exhibit small discrepancies when compared to those obtained under the MJP (obtained using the PMMH method described in 
\cite{GoliWilk15}). This is unsurprising given the discrete nature of the synthetic data. Nevertheless, posterior samples 
under the CLE are consistent with the true values that produced the data. Moreover, the inference scheme for the MJP gave 
a minimum ESS per second of 0.0062. Hence, for this example, sacrificing a small amount of posterior accuracy by using the CLE as an inferential model 
gives an increase in overall efficiency of a factor of over 3 orders of magnitude. Given the additional computational complexity of the 
remaining applications, in what follows we focus on either the CLE or Poisson leap as the inferential model. 

\subsection{Lotka--Volterra model}\label{lv}

The Lotka--Volterra system comprises two biochemical species (prey and predator) and three 
reaction channels (prey reproduction, prey death and predator reproduction, predator death). 
The reaction list is
\[
\mathcal{R}_1:\, \mathcal{X}_1 \xrightarrow{\phantom{a}c_1\phantom{a}} 2\mathcal{X}_1,\quad
\mathcal{R}_2:\, \mathcal{X}_1 + \mathcal{X}_2
\xrightarrow{\phantom{a}c_2\phantom{a}} 2\mathcal{X}_2 \quad \text{and}\quad
\mathcal{R}_3:\, \mathcal{X}_2 \xrightarrow{\phantom{a}c_3\phantom{a}} \emptyset .
\]
Let $X_{t}=(X_{1,t},X_{2,t})^T$ denote the system state at time $t$. 
The stoichiometry matrix associated with the system is
\[
S = \left(\begin{array}{rrr} 
1 & -1 & 0 \\
0 & 1 & -1 
\end{array}\right)
\]
and the associated hazard function is 
\[
h(X_{t},c) = (c_1 X_{1,t},c_2 X_{1,t}X_{2,t},c_3 X_{2,t})^T.
\] 
The CLE for this model is given by
\[
d\begin{pmatrix}
X_{1,t} \\
X_{2,t} 
\end{pmatrix} =
\begin{pmatrix}c_1 X_{1,t}-c_2 X_{1,t}X_{2,t} \\c_2 X_{1,t}X_{2,t}-c_3
  X_{2,t} 
\end{pmatrix}\,dt +
\begin{pmatrix}
  c_1 X_{1,t}+c_2 X_{1,t}X_{2,t}   & -c_2 X_{1,t}X_{2,t} \\ 
  -c_2 X_{1,t}X_{2,t} &c_2 X_{1,t}X_{2,t}+c_3 X_{2,t} 
\end{pmatrix}^{\frac{1}{2}}\,d
\begin{pmatrix}
W_{1,t} \\
W_{2,t}
\end{pmatrix},
\]
where $W_{1,t}$ and $W_{2,t}$ are independent standard Brownian motion processes.

\begin{figure}[t]
\centering
\psfragscanon
\psfrag{Density}[][][0.7][0]{Density}
\psfrag{log(c1)}[][][0.7][0]{$\log c_1$}
\psfrag{log(c2)}[][][0.7][0]{$\log c_2$}
\psfrag{log(c3)}[][][0.7][0]{$\log c_3$}
\includegraphics[width=0.34\textwidth,angle=-90]{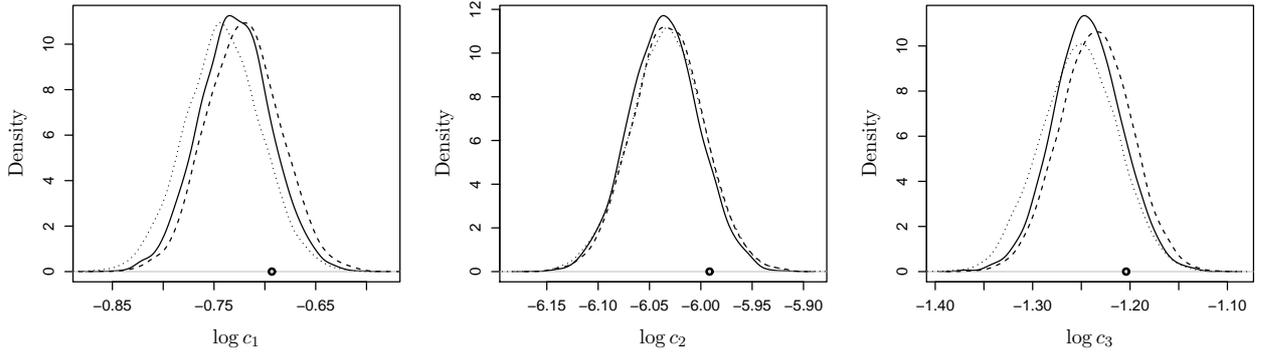}
\caption{Lotka--Volterra model. Marginal posterior distributions based on the output of CPMMH ($\rho=0.99$) using data sets $\mathcal{D}_{1}$ (solid lines), 
$\mathcal{D}_{2}$ (dashed lines) and $\mathcal{D}_{3}$ (dotted lines). The true values of $\log c_1$, $\log c_2$ and $\log c_3$ are indicated.}
\label{fig:figLV}
\end{figure}

We generated a single realisation of the jump process at 51 integer times via Gillespie's direct method 
with rate constants as in \cite{BWK08}, that is, $c=(0.5,0.0025,0.3)^T$ and an initial 
condition of $X_{0}=(100,100)^T$. We then obtained three data sets by corrupting the 
system state according to
\[
Y_{t}\sim \textrm{N}\left(X_t,\sigma^{2}I_{2}\right)
\]
where $I_{2}$ is the $2\times 2$ identity matrix and $\sigma\in\{1,5,10\}$ 
giving data sets designated as $\mathcal{D}_1$, $\mathcal{D}_2$ and $\mathcal{D}_3$ 
respectively. We took independent $N(0,10^2)$ priors for each $\log c_i$, $i=1,2,3$, and 
followed \cite{GoliWilk11} by setting $\Delta\tau=0.2$. The main monitoring runs consisted of 
$10^5$ iterations of MIS, CPMMH (with $\rho=0.99$) and PMMH. The results are summarised in 
Figure~\ref{fig:figLV} and Table~\ref{tab:tabLV}.

\begin{table}[t]
  \centering
  \small
  \begin{tabular}{@{}lllllll@{}}
    \toprule
    Data set & Algorithm  & $N$ & CPU(s) & mESS & mESS/s & Rel.  \\
    \midrule
             &           &             &                &      &       &  \\    
    $\mathcal{D}_{1}$ ($\sigma=1$)&\quad MIS   &\phantom{0}--&\phantom{0}14700&9218  &0.627  &13.5 \\
             &\quad CPMMH &\phantom{0}3 &\phantom{0}11280&8023  &0.711  &16.3 \\
             &\quad PMMH  &16           &\phantom{0}59730&2771  &0.046  &\phantom{0}1.0 \\
     \\  
$\mathcal{D}_{2}$ ($\sigma=5$) &\quad MIS   &\phantom{0}--&\phantom{0}14600&8139  &0.558  &14.3 \\
             &\quad CPMMH &\phantom{0}8 &\phantom{0}29780&3681  &0.124  &\phantom{0}3.2 \\
             &\quad PMMH  &20           &\phantom{0}75930&2959  &0.039  &\phantom{0}1.0 \\
             &           &             &                &      &       &  \\   
$\mathcal{D}_{3}$ ($\sigma=10$) &\quad MIS   &\phantom{0}--&\phantom{0}14690&6436  &0.438  &15.3 \\
             &\quad CPMMH &19           &\phantom{0}71520&3516  &0.049  &\phantom{0}1.7 \\
             &\quad PMMH  &28           &105770          &3031  &0.029  &\phantom{0}1.0 \\
    \bottomrule
  \end{tabular}
  \caption{Lotka--Volterra model. Number of particles $N$, CPU time (in seconds $s$), minimum ESS, minimum ESS per second and relative (to PMMH) minimum ESS per second. All results are based on $10^5$ iterations of each scheme.}\label{tab:tabLV}	
\end{table}

Figure~\ref{fig:figLV} shows that posterior samples are consistent with the true
values that produced the data, despite using an approximate inferential model
(the CLE). Table~\ref{tab:tabLV} shows a comparison of each competing inference
scheme. When using data set $\mathcal{D}_{1}$ ($\sigma=1$), CPMMH outperforms
PMMH by an order of magnitude (in terms of overall efficiency) and 
compares favourably with MIS. However, it is clear that as the measurement error
standard deviation ($\sigma$) increases, PMMH and CPMMH require more particles,
in order to effectively integrate over increasing uncertainty in the observation
process. Consequently, MIS outperforms PMMH and CPMMH when using
$\mathcal{D}_{2}$ ($\sigma=5$) and $\mathcal{D}_{3}$ ($\sigma=10$), although the
relative difference is less than an order of magnitude for MIS vs CPMMH. It is
worth noting that the rate of increase in $N$ is greater for CPMMH than for
PMMH. Increasing $\sigma$ appears to break down the correlation between
successive estimates of the log-posterior. Fixing the parameter values at the
posterior mean and estimating the correlation, denoted by $\rho_l$, between
$\hat{p}_{u}(y|c)$ and $\hat{p}_{u'}(y|c)$ gave $\rho_l=0.97$ for
$\mathcal{D}_1$, $\rho_l=0.91$ for $\mathcal{D}_2$ and $\rho_l=0.57$ for
$\mathcal{D}_3$. Nevertheless, we still observe a worthwhile increase in overall
efficiency of a factor of 2 for CPMMH vs PMMH, when using data set
$\mathcal{D}_{3}$ corresponding to the relatively extreme $\sigma=10$.

\subsection{Autoregulatory network}\label{ar}
In this section, we consider a simple autoregulatory network with two
species, $\mathcal{X}_{1}$ and $\mathcal{X}_{2}$ whose time-course behaviour
evolves according to the set of coupled reactions
\begin{align*}
\mathcal{R}_1&:\, \emptyset \xrightarrow{\phantom{a}c_1\phantom{a}} \mathcal{X}_1,\\
\mathcal{R}_2&:\, \emptyset \xrightarrow{\phantom{a}c_2\phantom{a}} \mathcal{X}_2,\\
\mathcal{R}_3&:\, \mathcal{X}_1 \xrightarrow{\phantom{a}c_3\phantom{a}} \emptyset,\\
\mathcal{R}_4&:\, \mathcal{X}_2 \xrightarrow{\phantom{a}c_4\phantom{a}} \emptyset,\\
\mathcal{R}_5&:\, \mathcal{X}_1 + \mathcal{X}_2 \xrightarrow{\phantom{a}c_5\phantom{a}} 2\mathcal{X}_2 .
\end{align*} 
Essentially, reactions $R_{1}$ and $R_{2}$ represent immigration and reactions $R_{3}$ and
$R_{4}$ represent death. The species interact via $R_{5}$. Let $X_{t}=(X_{1,t},X_{2,t})^T$ denote the system state at time $t$. 
The stoichiometry matrix associated with the system is
\[
S = \left(\begin{array}{rrrrr} 
1 & 0 & -1 & 0 & -1\\
0 & 1 & 0  &-1 &  1
\end{array}\right)
\]
and the associated hazard function is 
\[
h(X_{t},c) = (c_1,c_2,c_3 X_{1,t},c_4 X_{2,t},c_5 X_{1,t}X_{2,t})^T.
\] 

\begin{figure}[t]
\centering
\psfragscanon
\psfrag{t}[][][0.7][0]{$t$}
\psfrag{X1t}[][][0.7][0]{$X_{1,t}$}
\psfrag{X2t}[][][0.7][0]{$X_{2,t}$}
\includegraphics[width=0.4\textwidth,angle=-90]{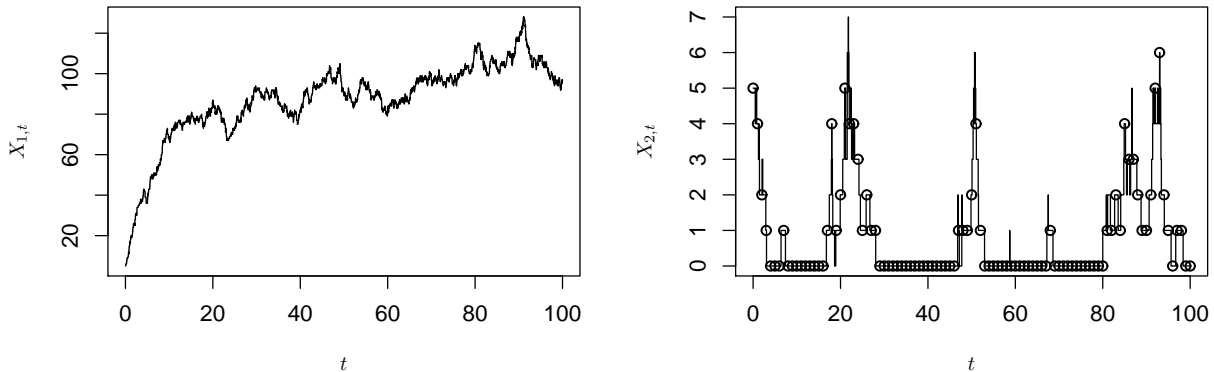}
\caption{Autoregulatory network. A single realisation of the jump process with $c=(10,0.1,0.1,0.7,0.008)^T$ and 
$X_{0}=(5,5)^T$. Observations are indicated by circles.}
\label{fig:auto}
\end{figure}

\begin{figure}[t]
\centering
\psfragscanon
\psfrag{Density}[][][0.7][0]{Density}
\psfrag{log(c1)}[][][0.7][0]{$\log c_1$}
\psfrag{log(c2)}[][][0.7][0]{$\log c_2$}
\psfrag{log(c3)}[][][0.7][0]{$\log c_3$}
\psfrag{log(c4)}[][][0.7][0]{$\log c_4$}
\psfrag{log(c5)}[][][0.7][0]{$\log c_5$}
\includegraphics[width=0.34\textwidth,angle=-90]{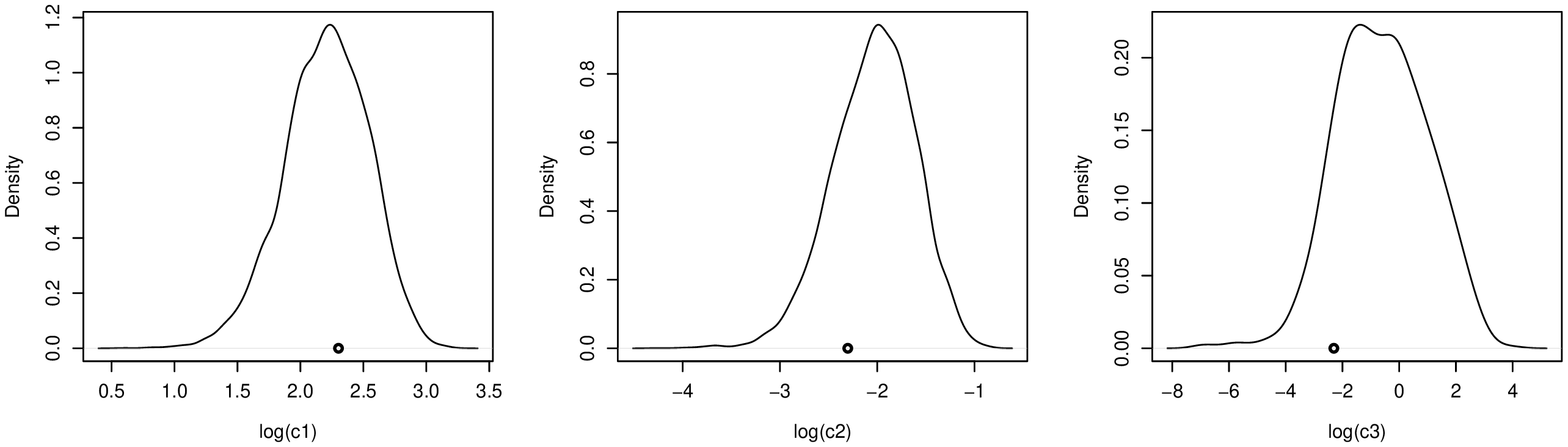}
\includegraphics[width=0.34\textwidth,angle=-90]{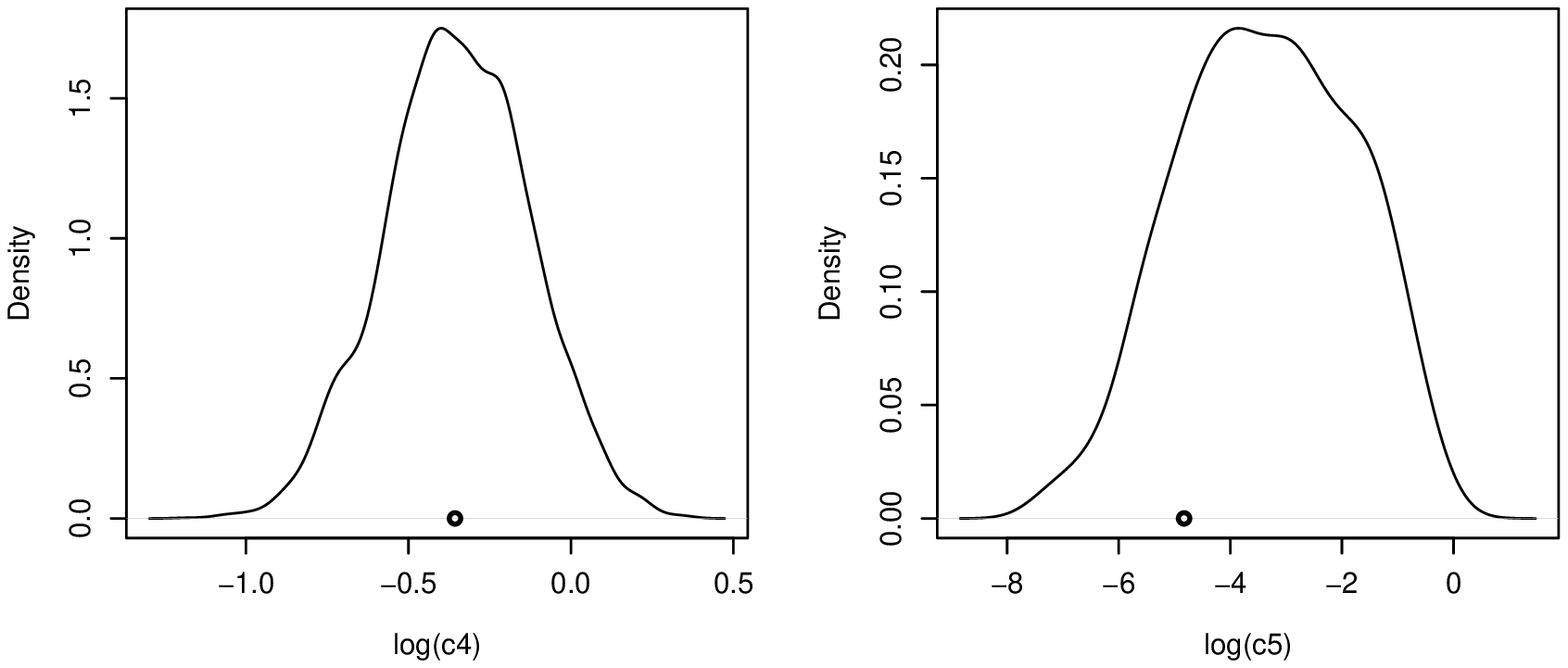}
\caption{Autoregulatory network. Marginal posterior distributions based on the output of CPMMH ($\rho=0.996$). 
The true values of $\log c_i$, $i=1,\ldots,5$, are indicated.}
\label{fig:figAR}
\end{figure}

We simulated a single realisation of the jump process at 101 integer times via Gillespie's direct method 
with rate constants $c=(10,0.1,0.1,0.7,0.008)^T$ and an initial 
condition of $X_{0}=(5,5)^T$. We then discarded the values of $X_{1,t}$ to leave observations of 
$X_{2,t}$ only. The full data trace used to generate the data set is given in Figure~\ref{fig:auto}. The 
inherently discrete nature of the data set coupled with long time periods where $X_{2,t}=0$ make 
applying the CLE impractical. We therefore use the Poisson leap approximation as the inferential model. 
To provide a challenging scenario, we assume error-free observation of $X_{2,t}$ so that 
step 2(d) of Algorithm~\ref{auxPF} assigns a weight of $0$ to the particle $x_t^i$ unless $x_{2,t}^i$ 
coincides with the observation at time $t$. We took a weakly informative Gamma$(10,1)$ prior for $c_1$ and Gamma$(0.1,0.1)$ priors for the 
remaining rate constants. We found little difference in sampled posterior values for a value of $\Delta\tau$ 
beyond $0.2$ and therefore used this value in our main monitoring runs which consisted of 
$10^5$ iterations of CPMMH (with $\rho=0.996$, which we found to work well for the partial observation scenario) 
and PMMH. We report results based on both the auxiliary and bootstrap 
particle filter driven pseudo-marginal schemes. The results are summarised in Table~\ref{tab:tabAR} and Figure~\ref{fig:figAR}.

\begin{table}[t]
  \centering
  \small
  \begin{tabular}{@{}llllll@{}}
    \toprule
    Algorithm  &   $N$ & CPU(s) & mESS & mESS/s & Rel.  \\
    \midrule
    CPMMH (APF)            &\phantom{0}20 &15580           &1272           &0.082 &6.2 \\
    PMMH\phantom{C} (APF)  &\phantom{0}55 &42010           &1302           &0.031 &2.4 \\
    PMMH\phantom{C} (BPF)  &200           &95800           &1263           &0.013 &\phantom{1.}1 \\
    \bottomrule
  \end{tabular}
  \caption{Autoregulatory network. Number of particles $N$, 
    CPU time (in seconds $s$), minimum ESS, minimum ESS per second and relative (to bootstrap filter driven PMMH) minimum ESS per second. All 
    results are based on $10^5$ iterations of each scheme.}\label{tab:tabAR}	
\end{table}

Again, we chose the number of particles $N$ by following the practical advice of \cite{tran2016} for CPMMH and \cite{sherlock2015} for 
PMMH. Inspection of Table~\ref{tab:tabAR} reveals that the bootstrap particle filter (BPF) driven PMMH scheme required $N=200$ particles. 
This reduces to $N=55$ when using the auxiliary particle filter (APF), and reduces further still to $N=20$ when strong and positive 
correlation is introduced between successive values of the random variables that drive the APF. Despite the APF driven scheme 
requiring many fewer particles than the BPF, overall efficiency (as measured by minimum ESS per second) is only increased by a factor 
of 2.4 due to the computational complexity of the conditioned hazard, which is used to propagate state particles within the APF. The 
correlated implementation gives a further increase of a factor of 2.6, giving a 6-fold increase in overall efficiency over the most basic 
PMMH scheme.

\subsection{Epidemic model}\label{epi}
\begin{figure}[t]
\centering
\psfragscanon
\psfrag{Density}[][][0.7][0]{Density}
\psfrag{c2}[][][0.7][0]{$c_2$}
\psfrag{c3}[][][0.7][0]{$c_3$}
\psfrag{c4}[][][0.7][0]{$c_4$}
\psfrag{c5}[][][0.7][0]{$c_5$}
\includegraphics[width=0.34\textwidth,angle=-90]{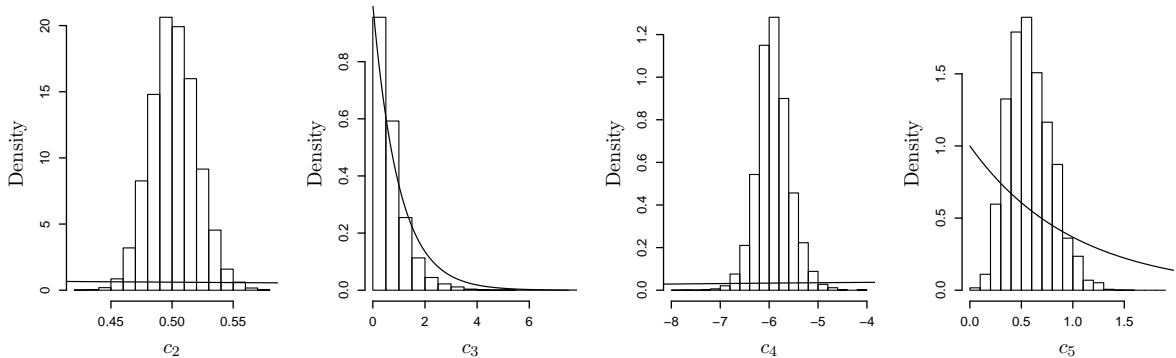}
\caption{Epidemic model. Marginal posterior distributions based on the output of CPMMH (histograms). Prior 
densities are given by the solid lines.}
\label{fig:figE}
\end{figure}

The Susceptible--Infected--Removed (SIR) epidemic model \citep[see][]{AnBr00} describes the evolution of two species 
(susceptibles $\mathcal{X}_{1}$ and infectives $\mathcal{X}_{2}$) via two reaction channels which correspond to an infection 
of a susceptible individual and a removal of an infective individual. The reaction equations are
\begin{align*}
\mathcal{R}_1&:\, \mathcal{X}_{1}+\mathcal{X}_{2} \xrightarrow{\phantom{a}c_{1}\phantom{a}}  2\mathcal{X}_{2}\\
\mathcal{R}_2&:\, \mathcal{X}_{2} \xrightarrow{\phantom{a}c_{2}\phantom{a}} \emptyset.
\end{align*}
The stoichiometry matrix is
\[
S = \left(\begin{array}{rr} 
-1 & 0\\
 1 & -1
\end{array}\right)
\]
and the associated hazard function is 
\[
h(X_t,c) = (c_{1} X_{1,t}X_{2,t}, c_{2} X_{2,t})^T.
\]
We consider a data set consisting of the daily number of pupils confined to bed
(out of a total of $763$) during an influenza outbreak in a boys' boarding school
in Great Britain, instigated by a single pupil. Hence, $X_{0}=(762,1)^T$. The data are displayed graphically in \cite{bmj1978} and converted
into counts in \cite{Fuchs_2013}. For completeness, we give the data in
Table~\ref{tab:tabB}. We work with the CLE which has the form
\begin{equation}\label{epi1}
  d\begin{pmatrix} X_{1,t}\\X_{2,t}\end{pmatrix}
  =\begin{pmatrix}
    -c_{1}X_{1,t}X_{2,t} \\
    c_{1}X_{1,t}X_{2,t}-c_{2}X_{2,t}
  \end{pmatrix}dt +
\begin{pmatrix}
  c_{1}X_{1,t}X_{2,t} & -c_{1}X_{1,t}X_{2,t} \\
  -c_{1}X_{1,t}X_{2,t}	 & c_{1}X_{1,t}X_{2,t}+c_{2}X_{2,t}
\end{pmatrix}^{1/2}d 
\begin{pmatrix} 
  W_{1,t}\\
  W_{2,t}
\end{pmatrix}.
\end{equation}
We further assume that the infection rate is a mean reverting diffusion process governed by the SDE
\begin{equation}\label{epi2}
d\log c_{1,t}=c_{3} (c_{4}-\log c_{1,t})dt+c_5 dW_{3,t}.
\end{equation} 
Hence, the inferential model is specified by (\ref{epi1}) and (\ref{epi2}), where $c_{1}$ is replaced by $c_{1,t}$ in (\ref{epi1}). We wish to 
infer $c=(c_{2},c_{3},c_{4},c_{5})^T$ based on measurements of $X_{2,t}$ only, giving a partially observed system. We took a normal N$(0,10^2)$ prior on the reversion level 
$c_{4}$ of $\log c_{1,t}$, and exponential Exp$(1)$ priors for the remaining parameters. For simplicity, we fixed the initial unobserved 
infection rate by taking $\log c_{1,0}=-6$. The discretisation level was fixed by taking $\Delta\tau=0.1$. The main monitoring runs consisted of 
$2\times 10^5$ iterations of CPMMH and PMMH. The results are summarised in Figure~\ref{fig:figE} and Table~\ref{tab:tabE}. It 
is evident that CPMMH outperforms PMMH in terms of overall efficiency (as measured here by minimum ESS per minute) by a factor of 7.

\begin{table*}[t]
  \centering
  \small
  \begin{tabular}{@{} lllllllllll@{}}
    
    Day               & 1 & 2 & 3 & \phantom{0}4  & \phantom{0}5   & \phantom{00}6   & \phantom{00}7   & \phantom{00}8   & \phantom{00}9  & \phantom{0}10 \\
    No. of infectives & 1 & 3 & 6 & 25 & 73  & 221 & 294 & 257 &236 & 189 \\ 
    \midrule
    Day               & \phantom{0}11  & 12 & 13 & 14 & 15 &  &  &  &  &    \\
    No. of infectives & 125 & 67 & 26 & 10 & \phantom{0}3  &  &  &  &  &    \\ 
    
  \end{tabular}
  \caption{Boarding school data.}\label{tab:tabB}
\end{table*}

\begin{table}[t]
  \centering
  \small
  \begin{tabular}{@{}llllll@{}}
    \toprule
    Algorithm  &   $N$ & CPU (m) & mESS & mESS/m & Rel.  \\
    \midrule
    CPMMH             &\phantom{0}90   & \phantom{0}2765&226 &0.08&7.2 \\
    PMMH\phantom{C}   &600             & 26338          &299 &0.01&\phantom{1.}1 \\
    \bottomrule
  \end{tabular}
  \caption{Epidemic model. Number of particles $N$, 
    CPU time (in minutes $m$), minimum ESS, minimum ESS per minute and relative minimum ESS per minute. All 
    results are based on $2\times 10^5$ iterations of each scheme.}\label{tab:tabE}	
\end{table}

\section{Discussion}\label{conc}

Exact (simulation-based) Bayesian inference for Markov jump processes (MJPs) is often rendered impracticable 
due to the requirement of many (millions of) exact simulations of the jump process. This 
computational cost can be controlled by replacing the inferential model with an approximation 
based on time-discretisation. Two such approximations that are routinely applied within 
the SKM literature are the (discretised) chemical Langevin equation (CLE) and Poisson leap. 
When using either approximation, the accuracy can be improved by introducing 
additional intermediate time-points between observation instances and integrating over the 
uncertainty associated with the induced latent process. As is the case for the MJP, the observed 
data likelihood under this implementation of time-discretisation remains intractable, requiring 
the use of PMMH. The key difference however, is that the number of intermediate time-points 
at which the latent process must be simulated can be pre-specified by the practitioner, 
with fewer time-points giving reduced computational cost, at the expense of accuracy of the 
inferential model. 

Taking either the (discretised) CLE or Poisson leap as the inferential model 
to be fitted, we increased the efficiency of PMMH by adapting the recently proposed correlated 
pseudo-marginal Metropolis--Hastings (CPMMH) algorithm \citep{deligiannidis2016,dahlin2015} to our setting. Positive 
correlation between successive observed data likelihood estimates was introduced by correlating the innovations that drive 
the proposal mechanism in the auxiliary particle filter. Essentially, the innovations are drawn from a kernel 
that satisfies detailed balance with respect to the innovation density. For a Gaussian innovation 
density (as is the case when using the CLE), a Crank--Nicolson proposal can be used. In the case of the 
Poisson leap, it is straightforward to map between Gaussian draws from a Crank--Nicolson proposal and the 
required Poisson variates. Whilst the degree of correlation present in the generation of the Gaussian innovations 
may be close to one, this does not necessarily directly translate into high correlations in the observed data 
likelihood. Nevertheless, in our experiments we see an improvement in performance relative to the standard 
PMMH scheme. 

For a fully observed, error-free immigration--death model, we found that 
it was possible to obtain an increase in overall efficiency (as measured by minimum effective 
sample size per second) of CPMMH over PMMH of around two orders of magnitude, whilst giving 
comparable performance to the modified innovation scheme of \cite{GoliWilk08}. To investigate the effect of measurement 
error, we applied each competing scheme to synthetic data generated from a Lotka--Volterra system 
and further corrupted with additive Gaussian noise. Not surprisingly, the performance of CPMMH worsens 
as the measurement error is increased, although we note that even in a relatively extreme scenario 
where the measurement error variance and average species values are of a similar order of magnitude, 
CPMMH outperforms PMMH by a factor of 2. We further applied CPMMH to a Poisson leap approximation of 
an autoregulatory network and to an SDE model of an influenza outbreak in a boys' 
boarding school. Despite only observing a subset of model components in both examples, 
we found that CPMMH outperforms PMMH by a factor of around 3 for the autoregulatory network and 
by a factor of 7 for the epidemic model. We note that bigger efficiency gains can be potentially 
achieved for extremely long data sets. For univariate models, it may be possible to scale the number of particles 
$N$ at rate $n^{1/2}$ (where $n$ is the number of observations) rather than at rate $n$, as 
is necessary for PMMH \citep{berard14}. For bivariate models, it may be possible to scale $N$ at rate $n^{2/3}$. 
See \cite{deligiannidis2016} for further discussion.  

The CPMMH algorithm can be improved upon in a number of ways. When using a particle filter to 
estimate the observed data likelihood, it may be beneficial to resample less often, thus 
preserving correlation between successive estimates of the observed data likelihood. Whether or 
not this is practically feasible will depend on the accuracy of the driving bridge proposal 
process. In scenarios with relatively few observations and when the proposal process is 
particularly effective, it may even be possible to avoid the resampling step altogether so 
that the particle filter is replaced by an importance sampler. The algorithm would also 
benefit from the availability of parallel computing architectures. In this case, the block 
pseudo-marginal method of \cite{choppala2016} could be used. A comparison of this approach 
with the methods described in this paper remains an area of active research. 

\

\noindent\textbf{Acknowledgements} The authors would like to thank the 
Associate Editor and two anonymous referees for their suggestions for
improving this paper.

\bibliographystyle{apalike}
\bibliography{bridgebib}

\appendix

\section{Modified innovation scheme}\label{mis}

We give a brief description of the modified innovation scheme (MIS) and refer the 
reader to \cite{whitaker2017a} and the references therein for further details. 

Consider the joint posterior of $c$ and the latent process $x$ under the CLE given by 
\[
\pi(c,x)\propto \pi_0(c)p(x|c)p(y|x)
\]
where $p(x|c)$ and $p(y|x)$ can be found in (\ref{xdens}) and (\ref{obsdens}). A Gibbs sampler 
can be used to generate draws from $\pi(c,x)$ by alternately sampling from the full conditionals
\begin{enumerate}
\item $p(x|c,y)$,
\item $p(c|x)$.
\end{enumerate}
It is straightforward to sample $p(x|c,y)$ using Metropolis within Gibbs coupled with a suitable 
blocking approach. For example, the latent process can be updated over each interval $[t-1,t+1]$, 
$t=1,2,\ldots,n-1$ with the modified diffusion bridge construct in (\ref{imp1}) used as the proposal 
mechanism. The use of overlapping blocks in this way ensures that latent process is updated at the 
observation times (as well as at all other intermediate times). The full conditional $p(c|x)$ can 
be sampled via Metropolis within Gibbs however for small values of $\Delta\tau$, dependence between 
the parameters and latent process can render this approach impractical. This well known problem 
is discussed at length in \cite{RobeStram01}. The issue is circumvented by the MIS via a reparameterisation. 
The basic idea is to draw parameter values conditional on a process whose quadratic variation does 
not determine $c$. For example, for a time interval $[0,T]$, conditioning on the innovations that drive the modified diffusion 
bridge construct leads to the continuous-time innovation process
\begin{align}
dZ_t&=\beta(X_t,c)^{-1/2}
\left(dX_t-\frac{x_{T}-X_t}{T-t}\,dt\right) \label{eqn:innovation}\\
&= \beta(X_t,c)^{-1/2}
\left\{\alpha(X_t,c)-\frac{x_{T}-X_t}{T-t}\right\}dt+dW_t \nonumber
\end{align}   
where $\alpha(X_t,c)=S\,h(X_t,c)$ and $\beta(X_t,c)=S\operatorname{diag}\{h(X_t,c)\}S^T$. A justification 
for conditioning on realisations of this process in a Gibbs sampler can be found in \cite{Fuchs_2013}. 
In practice, we work with a discretisation of (\ref{eqn:innovation}), that is, the modified diffusion 
bridge construct. For the induced invertible mapping $x=f(z)$ (where we have suppressed dependence of $f(\cdot)$ 
on $c$ and the values of the latent process at the observation times), the full conditional density required in step 2 is 
easily shown to be
\begin{equation}\label{eqn:innovationFC}
p(c|z)\propto\pi_{0}(c)p\{f(z)|c\}J\{f(z)|c\}
\end{equation}
where $p\{f(z)|c\}$ is given by (\ref{xdens}) and 
\[
J\{f(z)|c\}\propto \prod_{t=1}^{n-1}\prod_{k=1}^{m-1}\left|\beta(x_{\tau_{t,k-1}},c)\right|^{-1/2}
\]
is the Jacobian determinant of $f$. Naturally, the full conditional in (\ref{eqn:innovationFC}) will 
typically be intractable, requiring the use of Metropolis-within-Gibbs updates. We propose to 
update $\log c$ using random walk Metropolis with Gaussian innovations.

\end{document}